\documentclass[12pt]{article}
\usepackage{array}
\usepackage{amsmath}
\usepackage{amssymb}
\usepackage{amsfonts}
\usepackage{authblk}
\usepackage[dvips]{graphicx}
\usepackage{feynmf}

\numberwithin{equation}{section}

\input epsf
\def\beq{\begin{eqnarray}}
\def\eeq{\end{eqnarray}}

\def\lsim{\mathrel{\rlap{\lower3pt\hbox{\hskip0pt$\sim$}}
     \raise1pt\hbox{$<$}}}         
\def\gsim{\mathrel{\rlap{\lower4pt\hbox{\hskip1pt$\sim$}}
     \raise1pt\hbox{$>$}}}         

\newcommand{\pht}{\phantom}
\newcommand{\ud}{\mathrm{d}}

\newcommand{\km}{k_-}
\newcommand{\kp}{k_+}
\newcommand{\kpm}{k_\pm}

\begin{document}

\begin{flushright}
{ NYU-TH-05/11/21}
\end{flushright}
\vskip 0.9cm

\centerline{\Large \bf Resonance in Asymmetric Warped Geometry}
\vspace{0.3in}
\vskip 0.7cm
\centerline{\large Gregory Gabadadze, Luca Grisa and Yanwen Shang}
\vskip 0.3cm
\centerline{\em Center for Cosmology and Particle Physics}
\centerline{\em Department of Physics, New York University, New York, 
NY, 10003, USA}

\vskip 1.9cm


\begin{abstract}

We study the spectrum of an asymmetric warped braneworld model 
with different AdS curvatures on either side of the brane.
In addition  to the RS-like modes  we find  
a resonance state. Its mass is proportional to the geometric mean of the 
two AdS curvature scales, while the difference between them 
determines the strength of the resonance peak. There is a complementarity
between the RS zero-mode and the resonance: making the asymmetry stronger
weakens the zero-mode but strengthens the resonance, and vice versa.
We calculate numerically the braneworld gravitational potential 
and discuss the holographic correspondence for the asymmetric model.

\end{abstract}

\vspace{3cm}


\newpage

\section{Introduction}
In the RS2 model \cite{Randall:1999vf} a significant 
role is played by  the $\mathbb{Z}_2$ orbifold: 
The wavefunctions of  gravitational perturbation 
that are antisymmetric across the brane are 
projected out. On the other hand, one could consider more general geometries
in which the AdS curvatures on either side of the brane are different.
Such  branes can emerge  as consistent solitonic solutions 
in 5D $\mathcal{N}=1$ supergravities \cite{Cvetic:1992st}.  
These  domain walls interpolate between the AdS vacua with 
different cosmological 
constants, and  are stable as they saturate the BPS  bound.
Given the special properties of the RS spectrum, and its 
relevance for holography, it appears interesting to study  
similar issues in the braneworld with  the asymmetric bulk geometry.  
Some of the properties of the asymmetric models were discussed 
previously, see, e.g. \cite{Padilla},\cite{Koyama:2005br},\cite{Castillo-Felisola:2004eg}. In particular,
the present work was motivated by Ref.~\cite{Koyama:2005br}, 
which argued that the asymmetric model could provide UV completion to
brane induced gravity \cite{Dvali:2000rv}. If true, such a model could
be used to look at the strong coupling dynamics in \cite{Dvali:2000rv},
from a different perspective.
Below, we will give  
comprehensive studies of the spectrum and  
emphasize the similarities and differences 
of our results  with the earlier ones.

At energies that exceed  the  AdS curvature scales (but are 
still below the quantum gravity scale), both  the RS2 and asymmetric model  
are expected to have similar behavior. On the other hand,  
at  energies comparable to, or somewhat below of the AdS 
curvature scales, the asymmetric model  should behave differently from 
RS2.  Let us denote by  $k^2_\pm$ the two AdS curvatures, assuming for 
definiteness  $\km<\kp$.  At energies below  $\km$ one  can ``integrate 
out''  a slice of $AdS_+$ space. 
In the holographic description  this corresponds  to integrating out 
modes in the energy/momentum interval $\km <E< \kp$. 
The result is the RS2 model with AdS curvature $\km$ on both sides, 
in which, however, the orbifold projection is not imposed 
(i.e., the antisymmetric perturbations 
are kept). Moreover, in the low energy theory, 
while integrating out the slice of AdS, 
the brane-induced  Einstein-Hilbert (EH) term will appear. 
The coefficient in front of this term is going to be proportional 
to $(\kp^2-\km^2)/\kp^3$.  Hence, at energies  below 
$\km$, the original asymmetric model reduces to the RS2 model 
without the orbifold projection and with the brane-induced EH term,
similar to \cite{Dvali:2000rv}.

The properties of this hybrid low-energy model are well understood
\cite{Kiritsis:2002ca}. At energies above the scale 
 $m_*\sim \kp^3/(\kp^2-\km^2)$ gravity is similar to that of  
the DGP model \cite {Dvali:2000rv}. 
However, in the present construction this regime can 
not be reached, since the effective description in terms of the 
induced EH term is valid only at energies below   
$\km$, which is always smaller than $m_*$. Therefore, 
the DGP phase, emphasized in Ref. \cite {Koyama:2005br}, does not seem to 
be attainable in the asymmetric braneworld scenario
\footnote{It was claimed in \cite{Padilla} that the DGP phase can be reached if different
5D  Planck scales are assumed on two different sides of the  brane.
However, one could  perform asymmetric conformal transformations in the bulk to equate
the Planck scales  on both sides, at the expense of changing the nature of the junction condition
across the brane. In the present work we  will not consider such a possibility
since the  brane we keep in mind is a solitonic domain wall
solution of the type of Ref. \cite{Cvetic:1992st}, that arises
in a theory with a fixed 5D Planck scale. In principle,
the asymmetry in the 5D Planck scales could potentially be generated by an
asymmetric profile of a dilaton-type field.}.
Nevertheless, as we will describe below, the asymmetric models
exhibits a number of interesting  properties. The RS 
zero-mode is still 
present, as long as both AdS curvatures are nonzero. Moreover, 
because of the absence of the orbifold projection, both 
symmetric and antisymmetric modes survive. 
Conventionally, these modes can be grouped into two sets.
One set has properties similar 
to the RS spectrum \cite{Randall:1999vf}.
The net result of the  second set 
is a resonance state\footnote{A resonance in braneworlds was first 
discussed in Refs. \cite{grs}, in a different setup.}
with the mass $\sim \sqrt{\km \kp}$, 
and width $\sim \kp$. The strength of the resonance
peak is proportional to the difference of the the two AdS 
curvatures.  It vanishes in the limit when the two
curvatures are equal, and the corresponding set of states 
turns  into the antisymmetric modes that are
projected out from the RS model by the $\mathbb{Z}_2$ orbifold.

The paper is organized as follows.  In section 2, we 
set the action for the asymmetric model and discuss its general 
features. In section 3 we derive the complete spectrum of   
linearized perturbations and discuss some of its  
properties and implications. In section 4 we 
calculate the gravitational potential as seen by a 
braneworld observer in the asymmetric model and 
compare it with  the potential of the  RS2 model.
The RS model has been extensively studied from the point of view
of its holographic dual. It seems reasonable to expect that the 
asymmetric model also has a holographic description.  
We discuss these issues in  section 5. Conclusions are 
given in section 6.

\section{General formalism in the asymmetric setup}
The action of the braneworld with asymmetric warped
geometry is given by
\begin{equation}
\begin{split}
S=&\int\mathrm{d}^4x\int_{y>0}\!\!\!\!\mathrm{d}y\sqrt{-g}M_*^3\{2R+24\kp^2\}+
	\int\mathrm{d}^4x\int_{y<0}\!\!\!\!\mathrm{d}y
\sqrt{-g}M_*^3\{2R+24\km^2\}+\\
 &+\int_\mathrm{brane}\!\!\!\!\mathrm{d}^4x\sqrt{-g^*}
\{\tau+\mathcal{L}_\mathrm{SM}\}.
\end{split}
\end{equation}
The brane is located at $y=0$ where all the localized matter/gauge 
(called ``the Standard Model'')  fields live.
If the brane tension $\tau$ is fine-tuned to 
be $12M_*^3(\kp+\km)$, we have the
following solution for the metric in the bulk:
\begin{equation}
\begin{split}
\ud s^2=&e^{-2\kp |y|}\ud x^\mu \ud x_\mu-\ud y^2,
\quad\textrm{for}\, y>0, \, \textrm{and} \\
\ud s^2=&e^{-2\km |y|}\ud x^\mu \ud x_\mu-\ud y^2,
\quad\textrm{for}\, y<0.
\end{split}
\end{equation}
Indices $\mu, \nu$ run over directions along the brane.
One could think of this geometry as being two half-AdS spaces glued together 
across the brane at $y=0$.  $\kp^2$ and $\km^2$ are the $AdS$ curvatures on two
sides of the brane. In general they are different and
we define the ratio
\begin{equation}
\eta\equiv\frac{\km}{\kp},
\end{equation}
as a measure of the ``asymmetry'' of the model.

There are questions concerning the stability of the setup.
First, as we will show below, the small fluctuations in the model 
contain no tachyon  or ghost modes. 
Second, because of the asymmetry of the bulk 
cosmological constants, one may worry about 
the stability of the system with respect to the motion of 
the brane as a whole. However, when such a motion occurs, the 
gravitational background readjusts itself so that there is no net 
energy gain/loss on either side of the brane. Hence, there is no 
reason for such an instability. This is of course consistent with the 
brane being a bosonic part of a supersymmetric BPS object 
\cite{Cvetic:1992st}. The instability that could in general 
occur in the purely bosonic asymmetric case, 
is due to nonperturbative tunneling from one vacuum to the other
via bubble nucleation process.  However, 
this processes can be made exponentially small by appropriate choice 
of the parameters, and will be ignored below.

\begin{figure}[!htb]
\begin{center}
\includegraphics[width=0.4\textwidth]{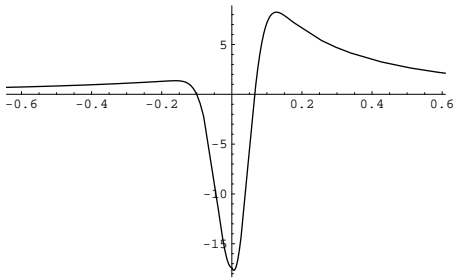}
\caption{\small\emph{Potential $V(z)$ where the $\delta$-function 
is regularized.}} 
\label{fig:potential}
\end{center}
\end{figure}

Following \cite{Randall:1999vf} we consider small gravitational
perturbations 
$G_{\mu\nu}=e^{-2\kpm|y|}(\eta_{\mu\nu}+h_{\pm\, \mu\nu}(x, y))$
in the gauge 
$\partial^\mu h_{\pm\,\mu\nu}=h^{\pht{\pm\,}\mu}_{\pm\,\mu}=0$.  
We can choose this gauge and keep the brane straight at $y=0$
in a source-free approximation. Performing the coordinate transformation
\begin{equation*}
z\equiv sgn(y)\left(e^{\kpm|y|}-1\right)/\kpm,
\end{equation*}
and field redefinition 
$h_{\pm\, \mu\nu}\equiv \psi^{(m)}_{\pm\, \mu\nu} e^{-|\kpm|y/2}e^{ipx}$, 
where $m^2\equiv p^2$, we find the ``Schr\"odinger'' equation for $\psi^{(m)}$.
Ignoring the tensorial structure it reads 
\begin{equation}
[-\frac{1}{2}\partial^2_z +V_\pm(z)]\psi_\pm^{(m)}(z)=m^2\psi_\pm^{(m)}(z),
\end{equation}
where the potential is given by
\begin{equation}
\label{equ:potential}
V_\pm (z)=\frac{15 k^2_\pm}{8(k_\pm |z|+1)^2}-\frac{3}{4}(k_+ +k_-)\delta(z).
\end{equation}
This potential is sketched in Fig.~\ref{fig:potential} 
with the $\delta$-function 
regularized.  It differs from that for the RS model as
the heights of the two barriers on different sides of the brane
are not equal.

We can identify four regimes for the potential: (1). Above both barriers, 
the gravity should behave as a $5$-D theory; (2). at energies
between the two peaks, one would expect novelties since
this regime is absent in RS2; gravitons of such energies can pass 
the barrier through tunneling on one side, but they are allowed to freely 
propagate on the other side as their energy is above the peak;
(3). Below both barriers we are in what could be called the  RS-phase, 
where we expect the model to exhibit properties similar to RS.
Furthermore, the zero-mode, if exists, is localized on the 
brane and gives rise to effective $4$-D interactions.

General solution of this equation can be found as
\begin{equation}
\psi^{(m)}_\pm(z)=(|z|+1/k_\pm)^{1/2}\left[
A_\pm Y_2(m(|z|+1/k_\pm))+ B_\pm J_2(m(|z|+1/k_\pm))\right].
\end{equation}
Here and below $Y_2$ and $J_2$ are Bessel functions and $A_\pm$ and $B_\pm$ are
constant coefficients to be determined by boundary conditions.  
The junction conditions across the brane are 
\begin{equation}
\label{eq:junction}
\psi^{(m)}_+(0)=\psi^{(m)}_-(0), \quad \textrm{and} \quad
\psi^{(m)}_+(0)'-\psi^{(m)}_-(0)'+\frac{3}{2}(k_+ + k_-)\psi^{(m)}(0)=0,
\end{equation}
where the derivative is with respect to $z$.

We normalize the wave functions in the following way.  While
$z\rightarrow \pm \infty$ all wave-functions $\psi^{(m)}_\pm (z)$
asymptote to plain waves.  It can be checked that this is 
equivalent to the usual orthonormal condition in the 
limit $L\rightarrow +\infty$, where
$L$ is the distance in the $z$-coordinate between the brane-world
and a fictitious regulator brane.  Since for $z\rightarrow \pm \infty$,
\begin{equation}
\sqrt{z}Y_2(mz)\sim \sqrt{\frac{2}{\pi m}}\sin(mz-\frac{5}{4}\pi),
\quad
\sqrt{z}J_2(mz)\sim \sqrt{\frac{2}{\pi m}}\cos(mz-\frac{5}{4}\pi),
\end{equation}
it is natural to parametrize the normalization coefficients
as:
\begin{equation}
\label{equ:generic.sol}
\psi^{(m)}_\pm(z)=\epsilon_\pm\sqrt{
\frac{\pi m(|z|+1/k_\pm)}{2(1+\alpha_\pm^2)}}
\left[\alpha_\pm Y_2(m(|z|+1/k_\pm)) +J_2(m(|z|+1/k_\pm))\right],
\end{equation}
with $\alpha_\pm$ to be determined by the junction conditions 
\eqref{eq:junction}.  Using the properties of the Bessel's functions, 
they are found to be
\begin{gather}
\label{eq:junction_wave}
\frac{\epsilon_+}{\sqrt{k_+(1+\alpha_+^2)}}
[\alpha_+ Y_{2+}+J_{2+}]
=\frac{\epsilon_-}{\sqrt{k_-(1+\alpha_-^2)}}
[\alpha_- Y_{2-}+J_{2-}], \\
\label{eq:junction_derivative}
\frac{\epsilon_+}{\sqrt{k_+(1+\alpha_+^2)}}
[\alpha_+ Y_{1+}+J_{1+}]
=-\frac{\epsilon_-}{\sqrt{k_-(1+\alpha_-^2)}}
[\alpha_- Y_{1-}+J_{1-}].
\end{gather}
Here $Y_{1, 2\, \pm}\equiv Y_{1, 2}(m/k_\pm)$ 
and $J_{1, 2\, \pm}\equiv J_{1, 2}(m/k_\pm)$.  

We have introduced the sign factors $\epsilon_\pm$ in the parametrizations
above.  They are functions of $m/\kpm$ and take the values of $\pm 1$ only.
\footnote{In solving the junction conditions one has to take
the square of the ratio of the two equations. Doing so the sign
becomes ill-defined.  To recover it, we must choose the sign-factors 
$\epsilon_{\pm}$ accordingly.}

In the limiting case when one of the $k$'s is $0$, the wave-functions
on the flat side turn into plane waves. Both
equations \eqref{eq:junction_wave} and \eqref{eq:junction_derivative} 
are simplified in this case.  Assuming that the ``$-$''-side is flat,
the  wave-functions are simply given by $\psi^{(m)}(z)=\sin(m z+\beta)$, 
with a phase shift angle $\beta$ to be determined by the junction conditions.
The latter, derived directly from \eqref{eq:junction}, are 
\begin{gather}
\label{eq:mink_junc_wave}
\sqrt{\frac{\pi m}{2k(1+\alpha^2)}}[\alpha Y_2(m/k)
+J_2(m/k)]=\sin\beta,\\
\label{eq:mink_junc_deri}
\sqrt{\frac{\pi m}{2k(1+\alpha^2)}}[\alpha Y_1(m/k)
+J_1(m/k)]=-\cos\beta.
\end{gather}
We have omitted all the $+$ signs for the quantities on the LHS of
the two equations above.
It is no longer necessary to
include the sign factor $\epsilon_+$  since it can always 
be absorbed into the redefinition of
$\beta$.

\section{Spectrum and its properties}
In the following section we will discuss in details the solutions
to the boundary conditions described above.  First we verify whether 
the zero-mode of the  RS model still exists in the 
asymmetric scenario.  As we will see this indeed is the
case, however, the zero-mode becomes weaker as one of the 
AdS curvatures decreases and disappears completely
from the spectrum if the bulk space-time on one side
is flat.  Regarding the massive KK modes, we will find that
alongside the RS-like modes, there exists a new solution in the
asymmetric scenario which behaves as a resonance.

\subsection{Zero-mode wave function}
\label{sec:zeromode}
\begin{figure}[!htb]
\begin{center}
\includegraphics[width=0.4\textwidth]{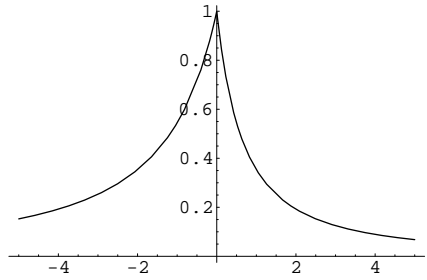}
\caption{\small\emph{Behavior of the zero-mode wave-function 
with $\km=.2$ and $\kp=1$.}}
\label{fig:zeromode} 
\end{center}
\end{figure}

In the RS2 model, the zero-mode
is localized on the brane and there it gives rise to the usual $4$-D
gravity at low energy.  In the asymmetric case,
the ``Schr\"odinger'' equation for the zero-mode reads 
\begin{equation}
[-\frac{1}{2}\partial^2_z +V(z)]\psi_0(z)=0,
\end{equation}
where $V(z)$ is the potential defined in the previous section
\eqref{equ:potential}. 
For generic $\kp$ and $\km$, the normalized solution to this 
equation can be found as
\begin{equation}
\psi_0(z)=\sqrt\frac{2\km\kp}{\km+\kp}\left[\frac{\theta(-z)}{(\km|z|+1)^{3/2}}
	+\frac{\theta(z)}{(\kp|z|+1)^{3/2}}\right],
\end{equation}
which is plotted in Fig.~\ref{fig:zeromode}.  Here $\theta(x)$
is the step function.  
This zero-mode is localized. The localization width is proportional 
to $1/\km$ on the left side and to $1/\kp$ on the right.   The prefactor 
of the wave-function is determined by fixing the norm of $\psi(z)$ to 
be $1$.  This sets the relation between
the bulk Planck mass $M_*$ and $4$-D Planck mass measured on the brane as 
\begin{equation}
M_{pl}^2=M_*^3 \left(\frac{1}{\kp}+\frac{1}{\km}\right).
\end{equation} 

Notice that as one of the two $AdS$ curvatures decreases, this zero-mode 
gradually delocalizes in the extra dimension while the effective 4-D
Newton coupling tends to zero. Eventually,
if the bulk space on one side of the brane is flat, the wave-function of
the zero-mode becomes non-normalizable.  Indeed as $\km\rightarrow0$ 
the overall multiplicative factor $[2\kp\km/(\kp+\km)]^{1/2}$
goes to zero.
Therefore in this limit the strength of the zero-mode vanishes on the 
brane. Equivalently the effective 4-D Planck mass becomes infinite, 
and Newtons constant tends to zero.

\subsection{Spectrum for $k_-=0$ and $k_+\ne 0$}
\label{sec:half_minkowski}
We now turn to the massive modes.  As a warm-up, we first discuss 
the spectrum in a simpler case where one of the curvatures 
vanishes so that the bulk space-time on one side of the brane is flat.
The boundary conditions are given by equations \eqref{eq:mink_junc_wave} and
\eqref{eq:mink_junc_deri}.   In the limit $m/k\ll 1$, their
solutions can be explicitly found.  In the leading order of 
the $m/k$ expansion
\begin{equation}
\begin{split}
Y_2(m/k)\sim -\frac{4 k^2}{\pi m^2}&, \quad 
Y_1(m/k)\sim -\frac{2 k}{\pi m}\qquad\textrm{and}\\
J_2(m/k)\sim \frac{m^2}{8 k^2}&, \quad 
J_1(m/k)\sim \frac{m}{2 k}. \quad 
\end{split}
\end{equation}
It is reasonable to expect that, in the limit $m/k\ll 1$,
the $Y$-terms in both equations dominate; this has to be verified on 
the solution by checking how the constant $\alpha$ defined in 
\eqref{equ:generic.sol} scales with $m/k$.

With these approximations we find
\begin{equation}
-\tan\beta=\frac{Y_2(m/k)}{Y_1(m/k)}=\frac{2k}{m},
\end{equation}
and, therefore, the strength of the KK modes on the brane is given by
\begin{equation}
\psi^{(m)}(0)^2=\sin^2\beta=\frac{1}{1+m^2/4k^2}.
\end{equation}
It is interesting to point out that this solution has a similar 
form as that of the DGP model \cite{Dvali:2000rv, Dvali:2001gx}, 
with the crossover distance $r_c=1/(2k)$.  This was first noticed by 
the authors of \cite{Koyama:2005br}.
We will discuss the physical significance of 
this similarity in subsection~\ref{sec:spectra_discussion}.

First we check that our approximation is consistent. Notice that 
in this approximation
\begin{equation}
\alpha \sim\sqrt{\frac{m}{k}}\frac{m}{k}.
\end{equation}
One can easily verify that in both equation \eqref{eq:mink_junc_wave}
and \eqref{eq:mink_junc_deri}, $Y$-terms do dominate as expected.  

\begin{figure}[!htb]
\begin{center}
\includegraphics[width=0.4\textwidth]{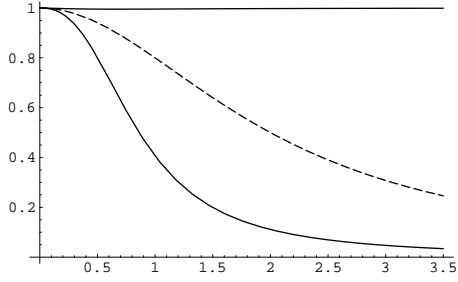}
\caption{\small\emph{\label{fig:solution} 
$\psi^{(m)}(0)^2\sim m/k$.  Dashed curve is for the DGP model.}}
\end{center}
\end{figure}

Exact solutions to \eqref{eq:mink_junc_wave} and 
\eqref{eq:mink_junc_deri} can be found by solving the 
quadratic equations.  The strengths of the KK modes on the brane 
are given as follows:
\begin{gather}
\Delta=\frac{2k}{\pi m}(Y_1^2+Y_2^2+J_1^2+J_2^2)-\frac{4 k^2}{\pi^2 m^2}
-(Y_1 J_2-Y_2 J_1)^2, \\
\alpha_{1, 2}=\frac{-(Y_1 J_1+Y_2 J_2)\pm\sqrt\Delta}
{Y_1^2+Y_2^2-2k/(\pi m)},\\
\psi^{(m)}_{1,2}(0)^2=\frac{\pi m}{2k(1+\alpha_{1, 2}^2)}
[\alpha_{1,2} Y_2+J_2]^2.
\end{gather}
We find that there exist two different solutions.
On Fig.~\ref{fig:solution} we plot both $\psi^{(m)}_{1, 2}(0)^2$ 
as functions of $m/k$.  Just for comparison we also plot the DGP 
spectrum with $r_c=1/(2 k)$.

\begin{figure}[!htb]
\begin{center}
\includegraphics[width=0.4\textwidth]{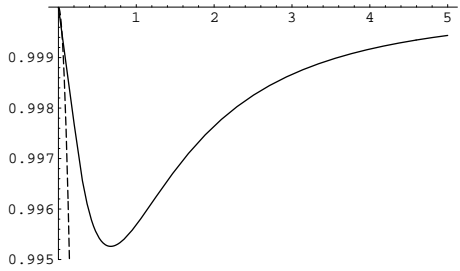}
\caption{\small\emph{\label{fig:solution_2} 
$\psi^{(m)}(0)^2\sim m/k$.  Small structure of the ``flat'' solution 
plotted alone.  Dashed curve is for the DGP model.}}
\end{center}
\end{figure}

As we see there exists a part of the 
spectrum that is almost constant.  This is of course expected.  
When we have a flat space on one side of the brane, KK gravitons 
of arbitrary energy can easily probe into the bulk 
\footnote{The ``flat'' spectrum is not exactly flat, 
but around masses of the order of the
AdS curvature it departs from being constant showing a tiny structure as shown
in Fig.~\ref{fig:solution_2}.  This is expected since at that mass scale 
the graviton mode can tunnel  into the curved side of the 
brane}.  However, in addition to the ``flat'' spectrum there exists 
a new solution which has a resonance 
type behavior.  The peak of this resonance is located at $m=0$
and its width is of the order of the curvature on the $AdS$ side.  

An immediate question following these observations arises. 
Does this new resonance state persist for arbitrary 
values of $\kp\ne\km$?  This is the issue that we will  
address in the next subsection.

\subsection{General spectrum for $\kp\km\ne 0$}
\label{sec:general.spectrum}
\begin{figure}[!htb]
\begin{center}
\includegraphics[width=0.4\textwidth]{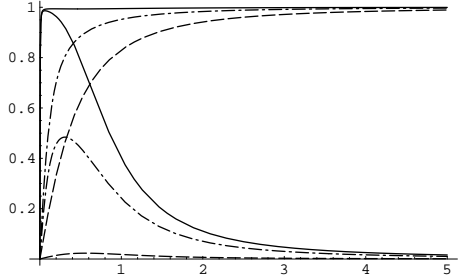}
\caption{\small\emph{Two solutions 
$\psi^{(m)}_{1, 2}(0)^2 \,\textrm{(Y-axis)}\, \sim m/k_+$,
with different values of $\eta=0.005$ (solid line), $0.2$ (dashed-dot line)
and $0.75$ (dashed line).}}
\label{fig:fsolutions}
\end{center}
\end{figure}

When both $k_\pm\ne 0$ and are not
equal to each other, one must solve equation \eqref{eq:junction_wave} and
\eqref{eq:junction_derivative} exactly.
We give below the square of the strengths
of the two solutions on the brane at $z=0$.  Defining
\begin{gather}
\eta\equiv\frac{k_-}{k_+},\\
\begin{split}
A=&\eta\,[(J_{2-}Y_{1+} + J_{1-}Y_{2+})(J_{1+}J_{2-}+ J_{2+}J_{1-})+\\ 
&+\!(Y_{1+}Y_{2-}+ Y_{1-}Y_{2+})(J_{1+}Y_{2-}+J_{2+}Y_{1-})\,],
\end{split}\\
B=\frac{2 \km}{\pi m}\cdot \sqrt{
\begin{split}
\eta[&(J_{2-}Y_{1+} + J_{1-}Y_{2+})^2+(J_{1+}J_{2-}+ J_{2+}J_{1-})^2+\\
&+(Y_{1+}Y_{2-}+ Y_{1-}Y_{2+})^2+(J_{1+}Y_{2-}+J_{2+}Y_{1-})^2]+\\
-&\eta^2[(J_{1+}Y_{2+}-J_{2+}Y_{1+})^2+4\kp^2/(\pi m)^2]
\end{split} }\qquad ,\\
D=-\eta[(Y_{1+}Y_{2-}+Y_{1-}Y_{2+})^2+(J_{2-}Y_{1+}+J_{1-}Y_{2+})^2]
+\left(\frac{2\km}{\pi m}\right)^2,
\end{gather}
on the brane, the two solutions are given by
\begin{gather}
\psi^{(m)}_1 (0)^2=\frac{\pi m}{k_+}\frac{[J_{2+} D+Y_{2+}(A-B)]^2}
{2[D^2+(A-B)^2]},\\
\psi^{(m)}_2 (0)^2=\frac{\pi m}{k_+}\frac{[J_{2+} D+Y_{2+}(A+B)]^2}
{2[D^2+(A+B)^2]}.
\end{gather}
The solutions with different values of $\eta$ are plotted on
Fig.~\ref{fig:fsolutions}.  There are always two distinct 
solutions.  One of them resembles the RS KK spectrum, 
and it will be referred to as the RS-like spectrum. 
The other one has a resonance type behavior and will be referred 
to as the resonance mode.  

Some comments on the qualitative properties of the two spectra are
in order.  First of all, when $\eta\rightarrow1$ the strength of the
resonance mode on the brane decreases and tends to 
zero.  This confirms the argument we have made in the introduction:
in the RS limit, the new modes we are considering becomes anti-symmetric 
wave-functions and automatically vanishes on the brane.  In the original 
RS model they are modded out by orbifolding the extra dimension. 
Only when we split $\kp$ from $\km$, we no longer have any symmetry 
reason to discriminate between the two solutions.
Neither of them is symmetric or anti-symmetric any more and we
must retain them both.

On the other hand, when $\eta\rightarrow 0, \textrm{or}\ +\infty$,
i.e., when one of the two sides is flat, we recover the results 
found in the previous subsection.  

It is a right place here to emphasize that the zero-mode 
strength is given by
\begin{equation}
|\psi^{0}(0)|=\sqrt{\frac{2\km\kp}{\km+\kp}}.
\end{equation}
When $\eta\rightarrow 0$ it vanishes while the 
resonance reaches its maximal strength.
On the other hand, the zero-mode has its maximal strength in the RS limit
when the resonance decouples.  We see that there is a complementarity 
between the resonance and zero-mode: when one disappears the 
other is present, and {\it vice versa}.  When $\km$ differs from $\kp$, 
the amplitude of the zero-mode weakens in favor of the resonance. 

\begin{figure}[!htb]
\begin{center}
\includegraphics[width=0.4\textwidth]{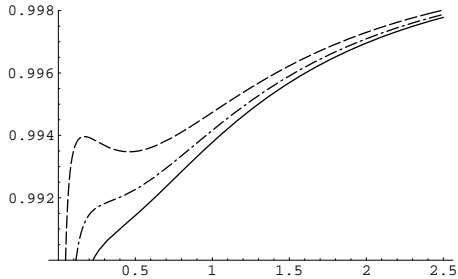}
\caption{\small\emph{$\psi_1^{(m)}(0)^2\,\textrm{(Y-axis)}\, \sim m/k_+$, with
$\eta=0.01$ (solid line), $0.008$ (dashed-dot line) and $0.005$ 
(dashed line) respectively.}}
\label{fig:detail_structure}
\end{center}
\end{figure}

Finally we point out that in the small $m/k_+$ region, 
when $\eta$ is sufficiently different from $1$, 
the once smooth RS-like spectrum starts developing a little distortion.
That is how, when $\eta$ changes, the spectrum deforms and links between the
smooth RS-spectrum ($\eta=1$) and the almost flat one ($\eta=0$), discussed
in the previous subsection.  Fig.~\ref{fig:detail_structure} demonstrates 
such a behavior.

It can be checked that when $m\ll\sqrt{\kp\km}$, 
both our solutions are approximately linear in $m$, in particular
\begin{equation}
|\psi_2^{(m)}|^2\sim \frac{\pi(1-\eta^{3/2})^2}{\eta(1+\eta)^2}m.
\end{equation}
On the other hand, in the limit $m\rightarrow \infty$, the RS mode 
approaches $1$ and the new mode tends to $0$ as $1/m^2$.

\subsection{On mixing and on resonance mass}
\label{sec:spectra_discussion}
First we turn to an important question: Are the
two distinct KK sectors orthogonal to each other?  If they are not, 
in the effective action on the $4$-D brane, a mixing term
of the two graviton modes would appear with a mixing coefficient $f_m$.

In the RS limit the two solutions are symmetric and anti-symmetric
wave-function respectively and are orthogonal.\\ 
To calculate the mixing
$f_m=\int\mathrm{d}z \psi_1^{(m)}(z) \psi_2^{(m)}(z)/\int\mathrm{d}z$,
we follow the normalization
procedure described in~\cite{Randall:1999vf}.  Introduce the 
regulator branes at $z_{c\pm}$ on both sides of the brane and 
correspondingly impose the new boundary condition there:
\begin{equation}
\partial_z \psi^{(m)}_\pm(z_{c\pm})=
-\frac{3\kpm}{2(\kpm z_{c\pm}+1)}\psi^{(m)}_\pm (z_{c\pm}).
\end{equation}
The zero-mode satisfies this condition and
remains unchanged.  When the regulator brane is at a finite distance from
the brane where matter is localized, these conditions will
quantize the KK masses in units of $1/z_{c\pm}$.
For simplicity we fix $z_{c+}=z_{c-}=z_c$ so that the discretized
mass spectrum is equally distributed.  In general, one could set both
$z_{c\pm}$ arbitrarily, but the result is a more complicated 
discretization of the mass spectrum of KK gravitons
since boundary conditions on both regulator branes must 
be satisfied simultaneously.  This, in principle, must give the same physical 
results when both regulator branes are taken to infinity but is mathematically 
more difficult to deal with.  When $z_c\rightarrow \infty$ the normalization 
constants of all the wave-functions are predominately those of plane 
waves, as in this limit the solutions we found approach free plane waves.
In this case it is easy to calculate the mixing $f_m$ which is given by
\begin{equation}
\begin{split}
f_m=&\epsilon_{1+}\epsilon_{2+}\cos\left(\arctan\alpha_{1+}
-\arctan\alpha_{2+}\right) \\
&+\epsilon_{1-}\epsilon_{2-}\cos\left(\arctan\alpha_{1-}
-\arctan\alpha_{2-}\right) \\
=&\frac{\epsilon_{1+}\epsilon_{2+}(1+\alpha_{1+}\alpha_{2+})}
{\sqrt{(1+\alpha_{1+}^2)(1+\alpha_{2+}^2)}}+
\frac{\epsilon_{1-}\epsilon_{2-}(1+\alpha_{1-}\alpha_{2-})}
{\sqrt{(1+\alpha_{1-}^2)(1+\alpha_{2-}^2)}}.
\end{split}
\end{equation}
Here $\epsilon_{1,2\pm}$ are the sign factors for each wave-function
as described before and $\alpha_{1,2\pm}$ are
the two solutions of equations \eqref{eq:junction_wave} and 
\eqref{eq:junction_derivative}.  We computed the expression  for the 
mixing coefficients numerically.  For any values of
the ratio $\eta$ and mass $m$, $f_m$'s  vanish with the 
extremely high accuracy ($10^{-13}$ or so). 
This suggests that the two spectra are indeed orthogonal 
to each other.

\begin{figure}[!htb]
\begin{center}
\includegraphics[width=0.4\textwidth]{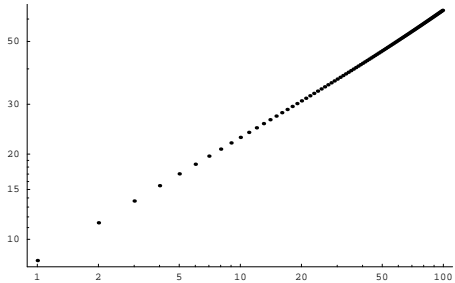}
\caption{\small\emph{Log$\sim$Log plot of the resonance mass 
$M$ versus $\km \kp$ (both axes are scaled by 100).}}
\label{fig:loglogplot_res_mass} 
\end{center}
\end{figure}

The next question of interest is the mass of the resonance mode,
defined as the location of its peak.
Unfortunately no analytic solution is known 
for the dependence of this mass $M$ on $\km$ and $\kp$. From
symmetry and dimensional analysis, it is reasonable 
to expect that $M\sim \sqrt{\km \kp}$.  This can be verified 
numerically as follows.  We solve for the mass $M$ 
numerically for different values of $\kpm$.  A log$\sim$log-plot 
of $M$ versus $\kp\km$ is shown in Fig.~\ref{fig:loglogplot_res_mass}.  
It suggests that $M\sim \sqrt{\kp \km}$.  
We point out that the width of the resonance is 
always of order of the greatest curvature ($\kp$).

Finally we would like to comment a bit more on the relation between 
the asymmetric model and the DGP model.  As we have seen,
when $\kp\sim \km$ the asymmetric and RS models are almost
identical and have nothing similar to DGP.  When
$\eta\ll 1$, we have the following three regions in terms of the 
distance $r$ as measured on the brane.  (1). $r>1/\kpm$: Both the asymmetric 
and RS model approach the same 4-D effective theory, in this regime 
the zero-mode dominates.  (2). $1/\kp\lesssim r\lesssim 1/\km$:
In this region we can integrate out the bulk space on the $\kp$ side and this
procedure leaves us with an induced $4$-D Einstein-Hilbert term in 
the effective action~\cite{Dvali:2001gx}.  Therefore we expect that the theory 
behaves as the DGP model with only a half of the bulk space.
In the extreme case when the bulk space on one side is flat,
we are in this region whenever $r\gtrsim1/k$, $k$ is the curvature 
of the $AdS$ side (it should be reminded that no zero mode is present
in this extreme case, therefore the previously discussed case does not apply).
This is perfectly consistent with what we 
discussed in Section~\ref{sec:half_minkowski}, where we
have shown that the spectrum agrees with that of the DGP model exactly for 
$r\gg 1/k$.  Notice, however, that this region coincides with the 
domain  $r\gg r_c$ 
in the corresponding DGP model where gravity already switches to the 
$5$-D behavior.  (3).  $r<1/\kpm$:  In this region the theory should give 
rise to $5$-D gravitational interactions.

\section{Gravitational potential on the brane}
It is instructive to compare the asymmetric model with $AdS$ curvatures $\kp$ 
and $\km$ with the RS model in which the curvature $k$ satisfies
\begin{equation}
\frac{1}{2}\left(\frac{1}{\kp}+\frac{1}{\km}\right)=\frac{1}{k}.
\end{equation}
As long as the above relation holds, the effective Planck mass $M_{pl}$
measured on the brane for both models is the same.

\begin{figure}[!htb]
\begin{center}
\includegraphics[width=0.5\textwidth]{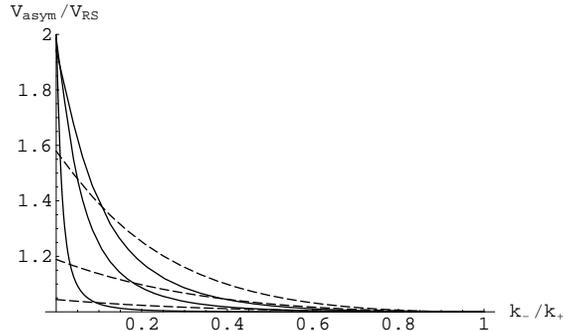}
\caption{\small\emph{$V_{asym}(r)/V_{RS}(r)$ versus
$\eta$.}}
\label{fig:KK_potential-eta} 
\end{center}
\end{figure}

Let us first look at the 2-particles gravitational potentials
on the brane-world.  They are given by 
\begin{equation}
V_{RS}(r)=\frac{1}{M_{pl}^2}\frac{m_1 m_2}{r}+\frac{1}{M_*^3} 
\int_0^\infty \ud m \frac{m_1 m_2 e^{-mr}}{r}\psi_{RS}(0)^{(m)\,2},
\end{equation}
in the RS, and 
\begin{equation}
\label{equ:newton}
V_{asym}(r)=\frac{1}{M_{pl}^2}\frac{m_1 m_2}{r}+
\frac{1}{M_*^3} \int_0^\infty \ud m \frac{m_1 m_2 e^{-mr}}{r}
[\psi_{1}(0)^{(m)\,2}
+\psi_{2}(0)^{(m)\,2}],
\end{equation}
in the asymmetric model.  It is hard to calculate the integrals 
analytically but numerical comparison
between the two model can be done.  Fig.~\ref{fig:KK_potential-eta} shows
the ratio $V_{asym}/V_{RS}$ versus $\eta=\km/\kp$.
Each curve represents $V_{asym}/V_{RS}$ measured at a fixed $r$.
For dashed curves, from bottom to up, $r\,\kp=25, 5, 1$ respectively,
and for solid curves, from top to down, $r\,\kp=1/5, 1/10, 1/50$.
As expected, when $\eta\rightarrow 1$, the two potentials are equal
at all distances. While for $\eta\ne1$ they are mostly different
at distances comparable to $1/\sqrt{\kp\km}\sim1/M_\mathrm{res}$.

A plot of $V_{asym}/V_{RS}$ versus $r\,k_+$ with a fixed value of $\eta=0.2$
is presented on Fig~\ref{fig:KK_potential-eta}. At very short distances, 
both models approach the usual $5$-D theory. At very 
large distances, the zero-mode in both models dominates and gives rise to
the same $4$-D gravitational potentials (we should stress that we have fixed
the effective 4-D Planck mass to be the same for both models).
The largest departure is at scales comparable to $1/M_\mathrm{res}$,
where the resonance is not exponentially
suppressed in the integral in \eqref{equ:newton}.
\begin{figure}[!htb]
\begin{center}
\includegraphics[width=0.5\textwidth]{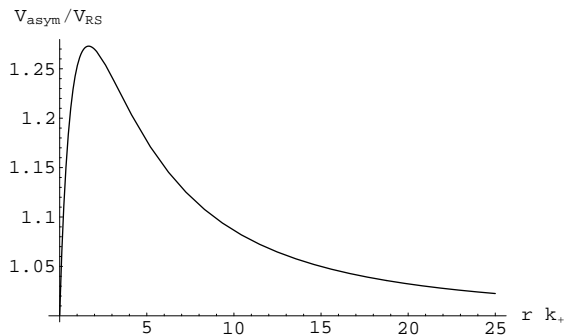}
\caption{\small\emph{$V_{asym}(r)/V_{RS}(r)$ versus
$r\,\kp$.}}
\label{fig:KK_potential-r} 
\end{center}
\end{figure}

\section{Comments on holographic correspondence}

The $AdS/CFT$ (holographic) correspondence
\cite{Maldacena:1997re,Witten:1998qj,Gubser:1998bc}
provides a powerful tool to explore  strongly coupled
field theories by looking at their weakly coupled gravity duals.
The holographic correspondence for  the RS2 model has been 
studied in great  detail. 
$AdS$ gravity is dual to a conformal field theory 
($CFT$) in lower dimension, and  the presence of the brane in the RS2 
model 
gives  rise to a UV cut-off and to  a massless 4D 
graviton on the $CFT$ side. 
The main evidence for the duality between RS 
and a UV cut-off $CFT$ coupled to 4D gravity were obtained 
in Refs. \cite{Gubser:1999vj,Duff:2000mt,Arkani-Hamed:2000ds}: 
The leading correction to the braneworld Newton potential 
found by RS on the gravity side
is reproduced by one-loop correction 
to the graviton propagator in $CFT$ \cite{Gubser:1999vj},
\cite{Duff:2000mt}. This radiative correction is evaluated from
the one-loop graviton self-energy  graph
where only the $CFT$ fields are running in the loop. For any 4D 
massless theory the finite part of such a loop 
is proportional to $\log(p_E^2)$, where 
$p_E^2$ is an euclidean  momentum square flowing through the loop 
diagram.  This term, after Fourier transformation, gives the $1/r^3$
correction to the Newton potential which exactly matches 
the correction due to the bulk KK gravitons \cite{Duff:2000mt}.
Furthermore, the very same calculation suggests 
the following dictionary: the RS2 model with bulk curvature 
$k^2$ is dual to $SU(N)$ $CFT$ with $N^2\sim M_\star^3/k^3$, where
$M_\star$ is the Planck mass of 5D gravity. The field theory 
UV cut-off $\Lambda$ is proportional to the curvature scale $k$. 

It is reasonable to expect that the holographic dual also 
exists for the asymmetric braneworld.  Schematically, 
the dual of the asymmetric warped braneworld 
would be a theory that consists of two sectors,  
$CFT_-\otimes CFT_+$, with number of colors
$N_\pm^2\sim M_\star^3/k_\pm^3$ and cut-off scales 
$\Lambda_\pm\sim k_\pm$ respectively. 
It is expected that the resonance is dual to a composite
state of $CFT_+\otimes CFT_-$, made of states living in both 
$CFT$'s. Its presence is due to the interaction terms, which 
have to be proportional  to the difference of the two curvatures, 
so that in the limit $\eta\rightarrow1$ the composite state 
disappears from the spectrum.

Assuming for definiteness that $\Lambda_+>\Lambda_-$, we can integrate 
out the modes with energies between $\Lambda_-$ and $\Lambda_+$
in the  $CFT_+$ sector.  The resulting low-energy effective field theory
reduces to $CFT_+\otimes CFT_-$ with the common UV cutoff $\Lambda_-$,
an additional induced cosmological constant 
(which has to be tuned to zero)  and, most importantly,  
the induced  EH term. The 
coefficient of the latter is  proportional to the 
difference between $\Lambda_+$ and $\Lambda_-$, and 
can be  evaluated by ``integrating out''  \cite{Arkani-Hamed:2000ds}, 
\cite {Redi}  a slab of the $AdS$ space between $1/k_+$ and $1/k_-$ 
\begin{equation}
2M_\star^3\int_{1/k_+}^{1/k_-}\mathrm{d}z\frac{1}{k_+^3z^3}
\int\mathrm{d}^4x
	\sqrt{-g^{(4)}}R^{(4)}=M_\star^3\frac{k_+^2-k_-^2}
{k_+^3}\int\mathrm{d}^4x
	\sqrt{-g^{(4)}}R^{(4)}.
\end{equation}
Except for this term, the resulting effective theory 
resembles the RS2 model. On the other hand,  
RS2 is dual to $CFT$, not to  $CFT\otimes CFT$. 
This difference is due to the orbifold projection imposed in RS, which 
results in the removal of the  antisymmetric  gravitational perturbation from 
the spectrum.  In the effective low energy 
theory of the asymmetric braneworld, no such a  
projection is imposed, and, therefore, both symmetric
and antisymmetric  modes are retained. Hence, the product structure 
${CFT_+}\otimes {CFT_ -} \to CFT_{sym}\otimes CFT_{asym}$ emerges. However, in 
the low-energy regime, 
only symmetric modes are coupled to  the brane localized 4D matter, 
while  the antisymmetric ones are null 
on the brane. Because of this, one of the two CFT's  
in the low energy  theory will not couple directly to the 4D matter. 
The effects of the high-scale coupling of this CFT to the 
localized matter
are encoded, in the low energy theory, in the induced 
EH term. This is consistent with the earlier observation that
the resonance is above the cutoff $\Lambda_-$.

\section{Conclusions}
We have studied in detail the spectrum of gravitational
perturbations in a brane-world scenario with asymmetric 
warped geometry.  We have shown that even when the two 
AdS curvatures $\kpm^2$  are different, 
the RS zero-mode still exists. Its strength on the brane is given by
$\sim [2\kp\km/(\kp+\km)]^{1/2}$, which vanishes if
one side of the brane becomes space-time flat.

A novelty  is found in the massive KK graviton 
spectrum\footnote{In Ref. \cite{Seahra:2005wk} quasi-normal states were
argued to exist in the standard RS spectrum (we thank Roy
Maartens for bringing this reference to our attention).
It would be interesting to see if similar states also appear
in the asymmetric part of the spectrum discussed in the present
work.}.  When the warp factors are different, there exists a 
resonance mode in addition to 
the RS-like spectrum. The peak of this resonance is located at
$M\sim\sqrt{\kp\km}$ and it has a width of the order of the
larger curvature scale. As the asymmetry between  $\kp\ne\km$ grows, 
the zero-mode becomes weaker and partially transmutes into the 
resonance mode.
In the limit when the bulk space-time on one side of the brane is flat, 
the resonance mode reaches its maximal strength while the 
zero-mode is no longer normalizable and disappears from the 
spectrum. On the gravity side we studied numerically 
the 2-body gravitational potential measured on the brane-world, and 
discussed the holographic description of the asymmetric model.
The obtained results can be useful for the particle physics model 
building via the AdS/CFT correspondence, as well as for 
the cosmology of early universe when studied in the context of 
warped space-times.

\vspace{0.4in}
{\bf Acknowledgments}
\vspace{0.2in}

We would like to thank Ofer Aharony, Elias Kiritsis and Michele Redi   
for useful discussions.  The work of GG was 
supported in part by  NASA Grant NNGG05GH34G, 
and in part by NSF Grant PHY-0403005. LG was supported by graduate 
student funds provided by New York University, and YS by Dean's 
Dissertation Fellowship.

\end{document}